\documentclass[11pt]{article}

\let\ssection=\section
\renewcommand{\section}{\setcounter{equation}{0}\ssection}

\def\p{{\partial}}

\newcommand{\br}{{\bf r}}

\newcommand{\bk}{{\bf k}}

\newcommand{\bB}{{\bf B}}
\newcommand{\vOmega}{{\mathbf{\Omega}}}

\def\det{{\rm det}}

\begin{document}

\setlength{\baselineskip}{16pt}

\title{Comment on ``Berry phase correction to electron density  in solids'' by Xiao et al.} 

\author{
C.~Duval\\
Centre de Physique Th\'eorique, CNRS, 
Luminy, Case 907\\ 
F-13288 Marseille Cedex 9 (France)\footnote{ 
UMR 6207 du CNRS associ\'ee aux 
Universit\'es d'Aix-Marseille I et II et Universit\'e du Sud Toulon-Var; Laboratoire 
affili\'e \`a la FRUMAM-FR2291.}
\\
Z.~Horv\'ath
\\
Institute for Theoretical Physics, E\"otv\"os
University\\
H-1117 BUDAPEST (Hungary)
\\
P.~A.~Horv\'athy
\\
Laboratoire de Math\'ematiques et de Physique Th\'eorique\\
Universit\'e de Tours\\
F-37 200 TOURS (France)
\\
L. Martina
\\
Dipartimento di Fisica dell'Universit\`a di Lecce
\\
I-73 100 LECCE (Italy)
\\ 
P.~C.~Stichel
\\
An der Krebskuhle 21\\
D-33 619 BIELEFELD (Germany)
}

\date{\today}

\maketitle

\begin{abstract}
The main result of  Xiao et al.  
 [Phys. Rev. Lett. 95, 137204 (2005)]
 follows from Hamiltonian mechanics.
\end{abstract}

\vskip0.5mm\noindent
\texttt{cond-mat/0509806},
Phys. Rev. Lett. {\bf 96}, 099701 (2006).


In a recent paper on the semiclassical dynamics of
a Bloch electron, Xiao, Shi and  Niu \cite{XSN} claim that, due to
a Berry curvature term,
 Liouville's theorem on the conservation of the
phase-space volume element would be violated, a fact which would have been overlooked so far. Then they suggest 
to restore invariance by including a pre-factor.

This Comment points out that no modifications of the existing theory are needed.

 Giving a Hamiltonian structure amounts indeed to giving a Hamiltonian and a Poisson-bracket which satisfies the Jacobi identity; the equations of motion read 
$\dot{\xi}=\{H;\xi\}$,  $\xi=(k_i,x^i)$  \cite{HamDyn}. 
The usual Hamilton equations, $\dot{\br}=\p_{\bf p}H$, $\dot{\bk}=-\p_{\bf r}H$, are 
only obtained when the coordinates $\bk, \br$ are canonical, 
$\{x^i,x^j\}=0, \{x^i,k_j\}=\delta^i_{\ j}, \{k_i,k_j\}=0$.

The system studied by Xiao et al.
{\it is}  Hamiltonian \cite{DHHMS}: 
their equations of motion \# (1a)-(1b) derive indeed from the Hamiltonian $H=\epsilon_n-eV$ and the Poisson brackets
\begin{eqnarray}
\{x^i,x^j\}&=\displaystyle\frac{\varepsilon^{ijk}\Omega_k}{1+e\bB\cdot\vOmega},
\label{xx}
\\[5pt]
\{x^i,k_j\}&=\displaystyle\frac{\delta^{i}_{\ j}+eB^i\Omega_j}{1+e\bB\cdot\vOmega},
\label{kk}
\\[5pt]
\{k_i,k_j\}&=-\displaystyle\frac{\varepsilon_{ijk}eB^k}{1+e\bB\cdot\vOmega}.
\label{xk}
\end{eqnarray}
The coordinates here are non-canonical, explaining the form of the equations of motion.

Let us emphasise that an ``abstract'' phase space carries {\it no natural} 
volume element~: the latter can only
be defined through a symplectic form $\omega_{\alpha\beta}$, which is 
the inverse of the Poisson-matrix 
$\omega^{\alpha\beta}=\{\xi^\alpha,\xi^\beta\}$.
Then the symplectic volume element is, in terms of {\it arbitrary}
coordinates on phase space,
\begin{equation}
dV=\sqrt{\det(\omega_{\alpha\beta})}\prod_{\alpha}d\xi^\alpha.
\label{volel}
\end{equation}
This is the usual $d\bk d\br$ only if the coordinates are canonical.

The general form of the Liouville theorem \cite{HamDyn} says that   
the symplectic volume element
{\it is} invariant w.r.t. the Hamiltonian flow. For the system considered, 
\begin{equation}
\sqrt{\det(\omega_{\alpha\beta})}=1+e\bB\cdot\vOmega,
\label{determinant}
\end{equation} 
precisely what Xiao et al. find.
Liouville's theorem is hence {\it not violated} and 
 the origin of the pre-factor is well understood.
It has  been used before, e. g., 
in \cite{DHH}, \# (3.1). 

The density of states  associated with a distribution function  
is $fdV$.  Putting $fdV\propto Dd\bk d\br$ yields,
together with Eqns. (\ref{volel}) and (\ref{determinant}) \cite{DHHMS}, 
the main result of Xiao et al., namely
$D\propto f(\bk,\br)\big(1+e\bB\cdot\vOmega\big)$. 

Later on, Xiao et al. study a planar system, their eqn. (17),
which corresponds to a planar charged particle in a magnetic field
and constant Berry curvature.
 This system
 is consistent \cite{DH} for any magnetic field, not only
for a uniform $B$, as Xiao et al. say. 
The critical case  $eB\theta=1$ is particularly interesting:
 det$(\omega_{\alpha\beta})=0$ and the system is singular. 
The symplectic (alias
Liouville) volume element, (\ref{volel}), vanishes \cite{DHH}. The Hamiltonian
reduction of the system leads to a simple planar system where all motions 
follow the Hall law,  whose quantization yields the ground states of the Fractional Quantum Hall Effect \cite{DH}. 

Acknowledgement.
We are indebted to J. Marsden for his encouragments.
P.H. and L.M. would like to thank  INFN and SINTESI grant by MIUR for financial
support.


\end{document}